\documentclass[10pt,pra,twocolumn,amssymb,nofootinbib,showpacs]{revtex4}
\usepackage{bm}
\usepackage{graphicx}
\usepackage{amsmath}
\begin{document}
\title{\bf Towards Quantum Superpositions of a Mirror:
an Exact Open Systems Analysis}\altaffiliation{Send correspondence
to: Stephen L. Adler, Institute for Advanced Study, Einstein
Drive, Princeton, NJ 08540 USA.  Phone: 609-734-8051, Fax:
609-924-8399, E-mail: adler@ias.edu}
\author{Angelo Bassi}
\affiliation{The Abdus Salam International Centre for Theoretical
Physics, Strada Costiera 11, 34014 Trieste,
Italy}\email{bassi@ictp.trieste.it} \affiliation{Istituto
Nazionale di Fisica Nucleare, sezione di Trieste, Strada Costiera
11, 34014 Trieste, Italy}

\author{Emiliano Ippoliti}
\affiliation{Department of Theoretical Physics, University of
Trieste, Strada Costiera 11, 34014 Trieste,
Italy}\email{ippoliti@ts.infn.it} \affiliation{Istituto Nazionale
di Fisica Nucleare, sezione di Trieste, Strada Costiera 11, 34014
Trieste, Italy}

\author{Stephen L. Adler}
\affiliation{Institute for Advanced Study, Einstein Drive,
Princeton, NJ 08540, USA}
\email{adler@ias.edu}

\begin{abstract}
We analyze the recently proposed mirror superposition experiment
of Marshall, Simon, Penrose, and Bouwmeester, assuming that the
mirror's dynamics contains a non--unitary term of the Lindblad
type proportional to $- [q,[q,\rho]]$, with $q$ the position
operator for the center of mass of the mirror, and $\rho$ the
statistical operator. We derive an exact formula for the fringe
visibility for this system. We discuss the consequences of our
result for tests of environmental decoherence and of collapse
models. In particular, we find that with the conventional
parameters for the CSL model of state vector collapse, maintenance
of coherence is expected to within an accuracy of at least 1 part
in $10^{8}$. Increasing the apparatus coupling to environmental
decoherence may lead to observable modifications of the fringe
visibility, with time dependence given by our exact result.
\end{abstract}
 \maketitle

There is currently much interest in experiments to create quantum
superposition states involving large numbers of particles, with
the ultimate aim of testing whether quantum superpositions of
macroscopic systems can be observed. Experiments performed to date
involve the superposition of oppositely circulating currents in
superconducting rings \cite{ref1}, and the diffraction of large
molecules \cite{ref2}, which give results in accord with quantum
mechanics.

Recently, Marshall {\it et al} \cite{ref4} proposed a novel
experiment which appears to be very promising, as the system to be
superposed can contain up to $10^{14}$ atoms, within present--day
technology. The experimental setup consists of an interferometer
for a single photon: in each arm of the interferometer there is a
cavity, and one of them (let us say, cavity A) has a movable
mirror mounted on a cantilever. Under suitable assumptions, the
Hamiltonian describing the composite system is \cite{ref7}:
\begin{equation} \label{eq1} H \; = \; \hbar
\omega_{c}( a^{\dagger}_{A}a^{\phantom \dagger}_{A} +
a^{\dagger}_{B}a^{\phantom \dagger}_{B}) \, + \, \hbar \omega_{m}
b^{\dagger}b \, - \, \hbar G a^{\dagger}_{A}a^{\phantom \dagger}_{A}(b
+ b^{\dagger});
\end{equation}
here $\omega_{c}$ is the frequency of the photon,
$a^{\dagger}_{A}$ and $a^{\dagger}_{B}$ are the creation operators
for the photon in the interferometer arms $A$ and $B$,
respectively, while $\omega_{m}$ and $b^{\dagger}$ are the
frequency and the phonon creation operator associated with the
motion of the center of mass of the mirror. The coupling constant
is $G=\omega_{c} \sigma/L$, where $L$ is the length of the cavity,
with $\sigma=(\hbar/2M \omega_m)^{1/2}$ the width of the mirror
wave packet and  $M$ the mass of the mirror.

A beam splitter places the photon in a state that is an equal
superposition of being in arm $A$ or $B$, so that the initial
state of the composite system is:
\begin{equation}\label{eq2}
|\psi_{0}\rangle \; = \;{1\over \sqrt{2}}\, \big[ |0\rangle_{A}
|1\rangle_{B} \; + \; |1\rangle_{A} |0\rangle_{B} \big]\,
|0\rangle_{m},
\end{equation}
with the mirror being at rest in its ground state.

Standard quantum mechanics predicts that at time $t$ the state vector
will be \begin{eqnarray}\label{eq3}
|\psi_{t}\rangle & = & e^{-{i \over \hbar} H t} |\psi_{0}\rangle
\; = \; {1\over \sqrt{2}}e^{-i \omega_c t} \, \left[ \frac{}{}
|0\rangle_{A} |1\rangle_{B}|0\rangle_{m} \right. \nonumber \\
& + & \left. e^{i \kappa^2(\omega_m t-\sin \omega_m t)} \;
|1\rangle_{A} |0\rangle_{B}|\alpha_t \rangle_{m} \right];
\end{eqnarray}
here we have written $\kappa =G/\omega_m$ and  $|\alpha_t
\rangle_{m} $ denotes a unit normalized mirror coherent state with
complex amplitude $\alpha_t =\kappa (1-e^{-i\omega_m t})$.
According to (\ref{eq3}), the state of the mirror changes into the
superposition of being at rest (when the photon travels through
arm B) and oscillating between $0$ and $4\kappa \sigma$ (when the
photon travels in arm A and hits the mirror); thus, in order to
create a spatially separated superposition of the two states $|0
\rangle_{m}$ and $|\alpha_t \rangle_{m}$, one has to require
$\kappa \geq 1/4$.

The physically interesting quantity of the experiment is the
visibility of the photon; this quantity is related to the
off--diagonal element of the reduced density matrix $\rho_{p} =
\makebox{Tr}_{m}[\rho]$ of the photon, where $\rho$ is the full
density matrix of the photon+mirror system. In our case, $\rho \;
= \; |\psi_{t}\rangle \langle \psi_{t}|$ so, after tracing over
the mirror states, and using the relation ${}_m \langle 0|\alpha_t
\rangle_m=e^{-{1\over 2}|\alpha_t|^2}$, the reduced density matrix
$\rho_{p}$ has as the coefficient of the off--diagonal term
$|1\rangle_A\, {}_A\langle 0| \otimes |0\rangle_B\, {}_B\langle 1|
$ the factor ${1\over 2}f$, with
\begin{equation}
\label{eq6} f=
e^{i\kappa^2(\omega_m t-\sin \omega_m t)} e^{-\kappa^2(1-\cos \omega_m
t)}.
\end{equation}
As expected, quantum mechanics predicts full coherence when the two
states $|0\rangle_{m}$ and $|\alpha_t \rangle_{m}$ of the mirror have
the same spatial position (this happens when $t = 2\pi n/\omega_{m}$,
$n \in {\mathbb N}$), while coherence is destroyed when $|\alpha_t \rangle_{m}$
moves away from the rest position.

Our aim in this Letter is to study how the system evolves when the
dynamics is not governed by the standard (unitary) Schr\"odinger
equation, but by the Lindblad--type dynamics:
\begin{eqnarray}\label{eq10}
d\rho\over dt & = & -{i\over \hbar} [H,\rho] -{1\over 2} \eta
[q,[q,\rho]] \nonumber \\
& = & -{i\over \hbar} [H,\rho] -{1\over 2} \eta \sigma^2
[b+b^{\dagger},[b+b^{\dagger},\rho]],
\end{eqnarray}
where $q=\sigma(b + b^{\dagger})$ is the position operator
associated with the center of mass of the mirror. The relevance of
an analysis of this kind lies in the fact that Eq. (\ref{eq10})
represents a cornerstone both for the theory of open quantum
systems \cite{dec} and for collapse models \cite{ref6}. In the
first case, it describes the reduced evolution of a quantum
particle interacting with a gas, by combining the free quantum
dynamics and the scattering with the particle flux. In the second
case, it represents the statistical evolution of a wavefunction
undergoing a spontaneous stochastic localization process in space.
Our problem is to solve the dynamics represented by Eq.
(\ref{eq10}), so as to calculate $f$.

First, two of us (A.B. and E.I.) set up the following linear
stochastic equation that unravels Eq. (\ref{eq10}):
\begin{equation} \label{lse}
d |\psi_{t}\rangle = \left[ -\frac{i}{\hbar} H dt + \sqrt{\eta}\,
q\, dW_{t} - \frac{\eta}{2} q^2 dt \right] |\psi_{t}\rangle,
\end{equation}
where $W_{t}$ is a standard Wiener process defined on a
probability space $(\Omega, {\mathcal F}, {\mathbb P})$. It is
easy to show that $f$ can be written as follows:
\begin{equation}
f = \int_{-\infty}^{+\infty} {\mathbb E}_{\mathbb P} \left[
\langle x | \phi^{A}_{t}\rangle_{m} {}_{m}\langle \phi^{B}_{t}|x
\rangle \right] dx,
\end{equation}
where ${\mathbb E}_{\mathbb P}[...]$ denotes the stochastic
average with respect to ${\mathbb P}$ and
$|\phi^{A}_{t}\rangle_{m}$ and $|\phi^{B}_{t}\rangle_{m}$ are two
wavefunctions for the mirror; they are solutions of Eq.
(\ref{lse}) when $H$ is replaced by the reduced Hamiltonians
$H^{A}$ and $H^{B}$ respectively, with $H^{A}$ the effective
mirror Hamiltonian acting when the photon passes through
interferometer arm A, and with $H^{B}$ the corresponding effective
mirror Hamiltonian acting when the photon passes through arm B:
\begin{equation}\label{eq12}
H^A \; = \; \hbar\omega_mb^{\dagger}b-G(b+b^{\dagger}) \qquad H^B
\; = \; \hbar\omega_mb^{\dagger}b.
\end{equation}
Thus the problem reduces to that of finding the two solutions of
Eq. (\ref{lse}), given the two initial conditions
$|\psi^{A}_{0}\rangle_{m} = |\psi^{B}_{0}\rangle_{m} =
|0\rangle_{m}$.

The other author (S.L.A.) then found a way to work directly from
Eq. (\ref{eq10}). Defining an off--diagonal density matrix
$\rho_{OD}$ acting in the mirror Hilbert subspace by $ {}_A\langle
1|{}_B \langle 0|  \rho |0\rangle_A|1\rangle_B ={1\over
2}\rho_{OD}$, so that the factor $f$ introduced above is ${\rm
Tr}_{m} \rho_{OD}$, one can project out from Eq. (\ref{eq10}) the
evolution equation for $\rho_{OD}$,
\begin{eqnarray}\label{eq11}
d \rho_{OD}(t) \over dt & = & -\frac{i}{\hbar}H^A\rho_{OD}(t)
+\frac{i}{\hbar}\rho_{OD}(t)H^B \nonumber \\
& & -{1\over 2} \eta \sigma^2
[b+b^{\dagger},[b+b^{\dagger},\rho_{OD}(t)]],
\end{eqnarray}
By a combined use of the interaction picture, the generalized
Baker--Hausdorff formula, and cyclic permutation under the trace
${\rm Tr}_{m}$, one gets directly ${\rm Tr}_{m} \rho_{OD}$ and
thus $f$.

Details of both methods of calculation have been given in
\cite{ref8}; we give here only the result, which is
\begin{eqnarray}\label{eq13}
f \; = \; \makebox{Tr}_m \rho_{OD}(t) & =
& e^{i\kappa^2(\omega_m t-\sin \omega_m t)-\kappa^2(1-\cos
\omega_m t)} \nonumber \\
& \times & e^{-{3\over 16} \eta \ell^2 \left(t-{4\over 3} {\sin
\omega_m t \over \omega_m } + {\sin 2 \omega_m t \over 6
\omega_m}\right)}, \qquad
\end{eqnarray}
where we have defined $\ell=4\kappa \sigma$ (so that $\ell$ is the
maximum mirror center of mass displacement in the state
$|\alpha_t\rangle$). The first term of $f$ in Eq. (\ref{eq13})
corresponds to the standard quantum result --- compare with Eq.
(\ref{eq6}) --- while the second term gives the correction due to
the non--unitary part of the evolution: as expected, it induces a
damping of the off--diagonal elements of $\rho_{p}$, which
increases in time.

The above result should be compared with the ``back of the
envelope'' calculation one can make by simply multiplying the
standard formula (\ref{eq6}) for $f$ by the factor
$e^{-\frac{1}{2} \eta \ell^2 t}$ coming from Eq. (\ref{eq10}) when
the unitary part of the evolution is omitted and one takes into
account the largest distance between the two states
$|0\rangle_{m}$ and $|\alpha_{t}\rangle_{m}$ of the mirror. This
heuristic expression is remarkably close to Eq. (\ref{eq13}). We
have learned that such a ``back of the envelope'' estimate has
been made earlier by Philip Pearle \cite{pcom} and applied to the
Marshall {\it et. al.} experiment, with results in agreement with
ours below. Our formula of Eq. (\ref{eq13}) goes beyond this
heuristic estimate by giving the exact time dependence of $f$.

We finally note that the exact solubility of the Marshall et. al.
model suggests that there may be further exact results associated
with the decoherent oscillator, and that this turns out to be the
case \cite{nref}. We now discuss two applications of our result.

\noindent {\it Application 1: test for collapse models.} There are
two principal proposals for modifications to quantum
superpositions of states with sufficiently large center of mass
displacement. Penrose \cite{ref5} has suggested that when two
macroscopic states are displaced so that their center of mass wave
packets no longer overlap, then new effects associated with
quantum gravity will destroy the coherent superposition. On the
other hand, there are the so called collapse models for
spontaneous stochastic state vector reduction \cite{ref6}. As
distinct from Penrose's proposal, collapse models require a much
larger center of mass displacement, typically taken as of order
$10^{-5}$ cm, for quantum superpositions to be destroyed.

Experiments performed to date fail by at least eleven orders of
magnitude \cite{ref3} to rule out the above proposals for
modifications to standard quantum theory. On the other hand, in
the proposed experiment of Marshall {\it et al} \cite{ref4} ---
developed to test Penrose's proposal --- the mirror center of mass
wave packet has a width $\sigma \sim 10^{-11}$ cm, and the two
states of the mirror that are superposed are displaced by at least
the wave packet width; in this circumstance, quantum mechanics
predicts full maintenance of coherence, whereas the Penrose's
proposal motivates a search for loss of coherence as evidenced by
reduced visibility of the interference fringes.

The quantity measured in the experiment is the maximum
interference visibility $\nu(t)$, which is equal to the magnitude
of $f$; thus, under standard quantum mechanical evolution of the
state --- see Eq. (\ref{eq6}) --- one has for the time dependence
of the visibility
\begin{equation}\label{eq7}
\nu(t)=e^{-\kappa^2(1-\cos \omega_m t)}.
\end{equation}
The strategy to test the macroscopic superposition of the mirror
then goes as follows. One measures the photon's visibility after
one period $T = 2\pi/\omega_{m}$ of the mirror's motion: if it is
close to 1, then no collapse of the mirror's wavefunction has
occurred; if on the contrary it is significantly smaller than 1, a
spontaneous collapse process is present which reduces the
superposition to one of its two terms. Of course, one must keep
control of all sources of environmental decoherence, which tend to
lower the observed visibility.

We now analyze the experiment within the context of collapse
models; more specifically, we consider the QMUPL model of wave
function collapse \cite{ref6}, which represents the small
displacement Taylor expansion of the GRW and CSL models
\cite{ref8}. This model is described by the following stochastic
differential equation
\begin{eqnarray}\label{eq8}
d\,|\psi_{t}\rangle & = & \left[ -{i \over \hbar}\, H\, dt +
\sqrt{\eta}\, (q - \langle q \rangle_{t})\, dW_{t} \right. \nonumber \\
& - & \left. {\eta\over 2}\, (q - \langle q \rangle_{t})^2 dt
\right] |\psi_{t}\rangle,
\end{eqnarray}
where $H$ is given by Eq.~(1), and $\langle q \rangle_{t} \equiv
\langle \psi_{t} | q | \psi_{t} \rangle$ is the quantum mechanical
expectation of the position operator $q$. Using the rules of the
It\^o calculus, the density matrix evolution corresponding to Eq.
(\ref{eq8}) is
\begin{equation}\label{eq9}
d\hat\rho=-{i\over \hbar} [H,\hat\rho] dt -{1\over 2} \eta
[q,[q,\hat\rho]] dt+ \sqrt{\eta} [\hat \rho,[\hat \rho,q]] dW_t.
\end{equation}

Since to observe interference fringes experimentally requires
passing an ensemble of identically prepared photons through the
apparatus, the relevant density matrix in the stochastic case is
the ensemble expectation $\rho=E[\hat \rho]$, which obeys Eq.
(\ref{eq10}). Accordingly, we can use Eq. (\ref{eq13}) to compute
the visibility, and we get \cite{ref10}:
\begin{eqnarray}\label{eq14}
\nu(t) & = & e^{-\kappa^2(1-\cos \omega_m t)}
\\
& \times & e^{-{3\over 16} \eta \ell^2 \left(t-{4\over 3}
{\sin \omega_m t \over \omega_m} + {\sin 2 \omega_m t \over 6
\omega_m}\right)}. \nonumber
\end{eqnarray}
This shows how the quantum mechanical visibility of Eq.
(\ref{eq7}) is modified by stochastic reduction in the QMUPL
model.

After one mirror oscillation has been completed, at time $T = 2
\pi/\omega_m$, the visibility predicted by Eq. (\ref{eq14}) is
damped by a factor $e^{-\Lambda}$, with
\begin{equation}\label{eq15}
\Lambda= (3/16) \eta \ell^2 (2\pi/\omega_m),
\end{equation}
while standard quantum mechanics (when no external sources of
noise are present) predicts it to be zero. We must now address the
question of the value of the stochasticity parameter $\eta$; we
focus our attention on the QMUPL, GRW, and CSL collapse models,
which give considerably different values for $\eta$.

In the GRW model, one has $\eta=N\eta_0$, with $N$ the number of
nucleons in the mirror and with $\eta_0$ obtained by comparison of
Eq. (\ref{eq10}) with the small displacement Taylor expansion of
Eq. (6.12) of Bassi and Ghirardi's review \cite{ref6}. In terms of
the parameters $\lambda=10^{-16} {\rm s}^{-1}$ and $\alpha=10^{10}
{\rm cm}^{-2}$ of GRW, this gives $\eta_0  = {1\over 2}\lambda
\alpha \sim 0.5 \times 10^{-2} {\rm s}^{-1} {\rm m}^{-2}$, which
for $N \sim 3 \times 10^{15}$ nucleons in the mirror, gives $\eta
\sim 10^{13} {\rm s}^{-1} {\rm m}^{-2}$. This is also the value
used in the QMUPL model.

In the CSL model, one can calculate $\eta$ from the small
displacement Taylor expansion of Eq. (8.23) of Bassi and Ghirardi
\cite{ref6}, details of which are given in \cite{ref8}. The result
of this calculation, in terms of the parameters $\gamma \sim
10^{-30} {\rm cm}^3 {\rm s}^{-1}$ and $\alpha=10^{10} {\rm
cm}^{-2}$ of CSL, together with the nucleon density $D=10^{24}
{\rm cm}^{-3}$ and the side length $S=10^{-3}{\rm cm}$ of the
cubical mirror, is
\begin{equation}\label{eq16}
\eta =  \gamma S^2 D^2 \left({\alpha \over \pi}\right)^{1\over 2}
\sim0.6 \times  10^{21} {\rm s}^{-1}{\rm m}^{-2}.
\end{equation}

Thus the CSL model gives an estimate of $\eta$ larger than that of
the QMUPL and QMSL models by a factor of $10^8$, and so we
continue the analysis using this more conservative value of
$\eta$. Taking $\kappa \sim 1/4$, so that the mirror excursion
$\ell$ is equal to its center of mass wave packet spread $\sigma
\sim 10^{-13}$ m, and with $2 \pi/\omega_m = 2 \times 10^{-3}{\rm
s}$, we get from Eqs. (\ref{eq15}) and  (\ref{eq16}) the result
$\Lambda \sim 0.2 \times 10^{-8}$, indicating that according to
the CSL model, coherence is maintained to an accuracy of better
than one part in $10^{8}$. Put another way, if the Marshall {\it
et al} experiment were to observe maintenance of coherence to 0.2
percent accuracy, it would set only the weak bound $\gamma<
10^{-24} {\rm cm}^3 {\rm s} ^{-1}$ on the CSL model stochasticity
parameter. This would be considerably better than the bound
$\gamma< 10^{-19} {\rm cm}^3 {\rm s} ^{-1}$  set \cite{ref3} by
fullerene diffraction experiments, but is still six orders of
magnitude away from a decisive test of the CSL model. Thus, while
providing a useful test of the Penrose model, the Marshall {\it et
al} experiment will leave very much open the question of whether
the laws of physics allow coherent superpositions of macroscopic
states.

\noindent {\it Application 2: test for environmental decoherence.}
We mention a second interesting application of our result, related
to decoherence tests. As previously discussed, the ultimate aim of
Marshall {\it et al} experiment is to test quantum superpositions
of macroscopic systems; in order to do this, one has to isolate
such a system from external noise, which tends to lower the
photon's visibility.

On the contrary, one can modify the experiment to deliberately
allow the mirror to interact with the environment, so as to
intentionally give a value of $\eta$ large enough to produce an
observable contribution from the second factor in  Eq.
(\ref{eq14}). The exact time dependence of Eqs. (\ref{eq13}) and
(\ref{eq14}) would then give a test of the environmental
decoherence model of Eq. (\ref{eq10}) and of the assumptions
\cite{dec} underlying this model.

The work of S.A. was supported in part by the Department of Energy
under Grant \#DE--FG02--90ER40542. The work of A.B. and E.I. was
supported in part by the Istituto Nazionale di Fisica Nucleare. We
wish to thank Dik Bouwmeester and Marco Genovese for helpful
conversations.


\begin{thebibliography}{99}

\bibitem{ref1}
C. H. van der Wal {\it et al}, Science {\bf 290}, 773 (2000); J.
R. Friedman {\it et al}, Nature (London) {\bf 406}, 43 (2000).

\bibitem{ref2}
W. Sh\"ollkopf and J. P. Toennies, {\it Science} {\bf 266}, 1345
(1994);  M. Arndt {\it et al}, Nature (London) {401}, 680 (1999);
O. Nairz {\it et al}, {\it J. Mod. Optics} {\bf 2000}, 2811; O.
Nairz {\it et al} {\it Phys. Rev. Lett.} {\bf 87}, 160401 (2001).

\bibitem{ref4}
W. Marshall {\it et al}, {\it Phys. Rev. Lett.} {\bf 91}, 130401
(2003).

\bibitem{ref7}
The Hamiltonian for a moving mirror interacting with an
electromagnetic field in a cavity has been studied by C.K. Law,
{\it Phys. Rev. A} {\bf 49}, 433 (1994); {\it Phys. Rev. A} {\bf
51}, 2537 (1995); S. Mancini, V.I. Man'ko and P. Tombesi, {\it
Phys. Rev. A} {\bf 55}, 3042 (1997); see also references therein.
The idea of using a moving mirror in a cavity to create
``macroscopic'' superpositions was first put forward by S. Bose,
K. Jacobs and P.L. Knight, {\it Phys. Rev. A} {\bf 59}, 3204
(1999). Ref \cite{ref4} contains the first proposal of an
interferometric setup involving a moving mirror, with the purpose
of testing ``macroscopic'' superpositions.

\bibitem{dec}
E. Joos and H. D. Zeh, {\it Z. Phys. B.} {\bf 59}, 223 (1985). D.
Giulini, E. Joos, C. Kiefer, J. Kupsch, I.--O. Stamatescu and H.D.
Zeh, {\it Decoherence and the Appearance of a Classical World in
Quantum Theory}, Springer, Berlin (1996). B. Vacchini, {\it Int.
J. Theor. Phys.}, to appear.

\bibitem{ref6}
The first consistent model of wave function collapse was developed
by G. C. Ghirardi, A. Rimini, and T. Weber, {\it Phys. Rev. D}
{\bf 34}, 470 (1986), and is known as the GRW or QMSL (Quantum
Mechanics with Spontaneous Localization) model. A continuous
stochastic version of this model, which also generalizes it to
systems of identical particles, has been developed by P. Pearle,
{\it Phys. Rev. A} {\bf 39}, 2277 (1989) and  G.C. Ghirardi, P.
Pearle, and A. Rimini, {\it Phys. Rev. A} {\bf 42}, 78 (1990), and
is known as the CSL (Continuous Spontaneous Localization) model. A
simpler continuous stochastic model for distinguishable particles
(QMUPL: Quantum Mechanics with Universal Position Localization)
has been proposed by L. Di\'osi, {\it Phys. Rev. A} {\bf 40}, 1165
(1989) and studied in detail by A. Bassi, {\it quant--ph/0410222}
(see also references therein). For recent reviews of these and
related models, and surveys of the literature, including
references to seminal papers by Di\'osi, Gisin, and Percival, see
A. Bassi and G. C. Ghirardi, {\it Phys. Reports} {\bf 379}, 257
(2003), and also S. L. Adler \cite{ref3}, Chapt. 6.

\bibitem{ref8}
S. L. Adler, A. Bassi, and E. Ippoliti, {\it quant--ph/0407084}.

\bibitem{pcom}
Private communication from P. Pearle to S. Adler.

\bibitem{nref}
S. L. Adler, {\it quant--ph/0411053}.

\bibitem{ref5}
R. Penrose, {\it Gen. Rel. Grav.} {\bf 28}, 581 (1996); R.
Penrose, in {\it Mathematical Physics 2000}, ed. by A. Fokas {\it
et al} (Imperial College , London, 2000).

\bibitem{ref3}
S. L. Adler, {\it Quantum Theory as an Emergent Phenomenon},
(Cambridge University Press, Cambridge, UK, 2004), Sec. 6.5.

\bibitem{ref10}
Subsequent to the derivation of Eq. (\ref{eq14}) by A.B. and E.I.,
we learned from a conference talk attended by S.A. that Dr. Ashok
Puri and collaborators are also studying decoherence effects on
the Marshall {\it et al} experiment. We do not have the details
needed to compare their method with ours.

\end{thebibliography}
\end{document}